\documentclass[12pt]{iopart}
\usepackage{iopams} 
	\usepackage{graphicx, xcolor}  

\begin{document}

\title[SOC and marginality in Ising spin glasses]{Self-organized critical behavior and marginality in Ising spin glasses}

\author{Auditya Sharma$^1$, Joonhyun Yeo$^2$ and M A Moore$^3$}

\address{$^1$ Department of Physics, Indian Institute of Science Education and Research, Bhopal, India}
\address{$^2$ Department of Physics,
Konkuk University, Seoul 143-701, Korea}
\address{$^3$ School of Physics and Astronomy, University of Manchester, Manchester M13 9PL, UK}


\begin{abstract}

We have studied numerically the states reached in a quench from various temperatures  in the  one-dimensional  fully-connected Kotliar, Anderson and Stein  Ising spin glass model. This is a model where there are long-range interactions between the spins which falls off as a power $\sigma$ of their separation. We have made a detailed study in particular of the energies  of the states reached in a  quench from infinite temperature and their  overlaps, including the spin glass susceptibility. In the  regime where $\sigma \le 1/2$, where the model is similar to the Sherrington-Kirkpatrick model, we find that the spin glass susceptibility diverges logarithmically with increasing $N$, the number of spins in the system, whereas for $\sigma> 1/2$ it remains finite. We attribute the behavior for $\sigma  \le 1/2$ to \emph {self-organized critical behavior}, where the system after the quench is close to the transition between states which have trivial overlaps and those with the non-trivial overlaps associated with replica symmetry breaking. We have also found by studying the distribution of local fields that  the states reached in the quench have marginal stability but only  when  $\sigma \le 1/2$.
\end{abstract}
\noindent{\it Keywords\/}: spin glasses, self-organized criticality, marginal stability

\submitto{\JSTAT}
\maketitle

\section{Introduction}

In this  paper we have  studied three topics related  to deterministic
quenches in a spin glass system. The  spin glass system is that of the
one-dimensional  long-range Ising  Hamiltonian introduced  by Kotliar,
Anderson and Stein (KAS) \cite{KAS}, which serves as a proxy model for
the     short-range     $d$-dimensional     Edwards-Anderson     model
\cite{edwards:75}. The  KAS model  has long-range  interactions between
the spins  which fall off  with a  power $\sigma$ of  their separation
distance. In  a quench  we start  from an initial  state, such  as the
fully  equilibriated state  at  a  temperature $T$  and  then apply  a
deterministic  algorithm,  such  as   the  ``greedy'',  ``polite''  or
sequential  algorithm \cite{parisi:95}  until  a state  is reached  in
which the energy  cannot be lowered further by flipping  just a single
spin. Much of our investigation has been of the case where the initial
state is at infinite temperature so that spins are randomly $\pm 1$ with  the quench being performed 
with the sequential algorithm.

The first study is  of the nature of the state  reached in the quench,
as revealed  by the form  of the  Parisi overlap function  $P(q)$. Our
finding  here is  that its  form is  determined by  the nature  of the
initial state  at temperature  $T$. If  $T> T_c$,  where $T_c$  is the
equilibrium transition temperature of the  spin glass system, then the
final  state  has  the  trivial overlap  of  the  paramagnetic  state,
$P(q)=\delta(q)$. When  $T< T_c$ its  form after the  quench resembles
that of the initial state. We  conclude that in a deterministic quench,
the  form of  the initial  state is  imprinted onto  the final  quenched
state.

The second study which we make is of the distribution of local fields
$p(h)$  in the  quenched state.  We review  the argument  of Anderson,
reported in  Ref.~\cite{palmer:79},  for the form  of $p(h)$  at small
fields and find  that our numerical data is consistent  with the state
generated  in the  quench  having \textit{marginal}  stability in  the
regime where mean-field  applies, that is for $\sigma  \le 1/2$, but that
outside this regime the quenched state is not marginal. The mean-field
limit   includes   the   Sherrington-Kirkpatrick  (SK)   model   which
corresponds  to the  case $\sigma  = 0$.  The form  of $p(h)$  after a
quench has been  much studied for the Ising  SK model \cite{parisi:95,
  eastham:06, horner:07,yan:15}.

The third  topic studied is  that of Self-Organized  Criticality (SOC)
which is the phenomenon where some large dissipative systems can be in
a scale-invariant critical state but without any parameter being tuned
to a critical value \cite{schenk:02}. It is believed that it is behind
the  fractal   features  \cite{mandelbrot:83}  associated   with  many
phenomena, such as  earthquakes, the meandering of sea  coasts and the
structure  of  galactic  clusters.   For  equilibrium  systems,  scale
invariant behavior is usually only found at critical points where some
parameter e.g.\ temperature is at  its critical value  $T_c$. However,
over the years a number of examples have been found of SOC behavior in
models  which have  very artificial  dynamical  rules such  as in  the
sandpile   model    \cite{bak:87}   and   the   forest    fire   model
\cite{drossel:92}. More  recently, Andresen  et al.~\cite{andresen:13}
have found SOC  features in the Sherrington-Kirkpatrick  (SK) model of
Ising  spin   glasses  which   were  absent  in   the  $d$-dimensional
Edwards-Anderson (EA) spin glass models. The signature of SOC behavior
for them was the size of the spin avalanches following a change in the
applied field; only when there was  a diverging number of neighbors as
in the  SK model  were the  avalanche sizes limited  by the  number of
spins $N$ in the system. 

Most studies of SOC behavior focus on {\it dynamical} features such as
the            size            of           avalanches            etc.\
\cite{Pazmandi:99,andresen:13,sharma2014avalanches}. In  this paper we
have studied  entirely {\it static}  features: in particular  the spin
glass  susceptibility  calculated via  the  overlaps  of the  quenched
states obtained from different initial  states.  We have found that it
diverges logarithmically with  the number of spins $N$  in the system,
provided that  the exponent $\sigma  \le 1/2$.  We thus  conclude that
when $\sigma  \le 1/2$  the system  reaches a  set of  quenched states
which  are  close to  a  critical  energy.  When  $\sigma >  1/2$  the
divergence of the susceptibility goes away, indicating that the quench
does  not then  take the  system to  a critical  state. It  is thought
\cite{mori:11} that systems  with $\sigma \le 1/2$ behave  just as the
SK    model.   Our    results    therefore    complement   those    in
Ref.~\cite{andresen:13}, where  they found that SOC  behavior was only
present for  the SK model, but  was lacking in the  $d$-dimensional EA
models, which correspond to values of  $\sigma > 1/2$ in the KAS model
\cite{ky:2003,katzgraber:09b,leuzzi:09,banos:12b,aspelmeier:16}. Furthermore,
Gon\c{c}alves  and  Boettcher  \cite{Boettcher:08}  studied  avalanche
sizes as  a function of $\sigma$  in the KAS model  and concluded that
$\sigma  =1/2$  was  indeed  the  borderline  value  above  which  the
avalanches changed their behavior as a function of system size $N$. An
extensive  study of  avalanches  in the  SK model  itself  is in  Refs.~\cite{Pazmandi:99,zhu:14}.

 Our interest in  the static aspects of SOC behavior  was triggered by
 our previous studies \cite{Sharma16}  of \textit{vector} spin glasses
 in the Sherrington-Kirkpatrick  (SK) model. We found  that the quench
 in those models reached metastable minima whose energy per spin $E_c$
 was  very close  to that  calculated for  the energy  which separates
 minima with zero overlap with each other from those which have a full
 replica  symmetry overlap  with each  other \cite{bm1981}.   In other
 words  the  quench takes  one  close  to  the critical  energy  which
 separates states with a trivial  $P(q)$ from those with a non-trivial
 $P(q)$. The  same type of  mean-field calculation fails in  the Ising
 case as  the states  reached in  a quench are  quite atypical  of the
 set of all  the  metastable   states  of   energy  $E$.    In  the
 thermodynamic limit the energy per spin reached after the quench from
 a random  initial state tends  to a well-defined limit,  dependent on
 the method used to flip the spins (e.g. \lq polite' or \lq greedy' or
 \lq sequential' algorithm etc.\ \cite{parisi:95}).  These observations
 are  consistent  with the  rigorous  arguments  of Newman  and  Stein
 \cite{newman:99}.   The   states  reached   in  the  quench   have  a
 distribution $p(h)$  of their local  fields $h_i= \sum_j  J_{ij} S_j$
 with interactions $J_{ij}$ among the  spins $S_i$, which is linear in
 $h$ at small fields, whereas for the totality of metastable states of
 energy $E$, $p(0)$ is finite \cite{roberts:81}.

Our  main finding  is that  for Ising  spin glasses  in the  SK region
$\sigma \le 1/2$,  the energy of the system after  the quench is close
to the critical energy $E_c$  which separates the metastable states of
the kind produced in the quench  which have no overlap with each other
from those  which would exist  at lower  energy which would  have full
replica symmetry  breaking overlaps. That  is Ising spin  glasses with
$\sigma \le 1/2$ behave very  similarly to vector spin  glasses, except
that for  Ising spin glasses the  definition of $E_c$ is  not that for
the set of  all states of energy $E$  as for the case  of vector spin
glasses  but  instead it  is  the  critical  energy for  those  states
produced  in the  quench (which  have a  distribution of  local fields
$p(h) \sim  h$ at small  $h$). In Ref.~\cite{andresen:13} \textit{the
  nature  of the  ordering associated  with the  SOC behavior  was not
  specified}.   If the  quench is  close to  this critical  energy one
would expect  there to be  a divergent spin glass  susceptibility; the
definition and  the study of  this susceptibility  is one of  the main
topics of  this paper. It  is a purely  static quantity: the  study of
avalanches  alone does  not provide  insights into  the nature  of the
incipient ordering associated with the SOC.

 There  is an  important  distinction between  Ising  and vector  spin
 glasses.   Edwards hypothesized  (for a  review see  \cite{baule:16})
 that systems like  powders or sand piles etc.\ could  be understood not
 by solving the full dynamics of  the system from its initial state to
 its final  resting state (which  is hard) but instead  by determining
 for  these systems  the  analogue of  the number  of  states in  spin
 glasses in which the spins are parallel to their local fields, (which
 is easy) \cite{bm1981}.  For vector spin glasses in the SK limit, his
 hypothesis has  utility.  It fails completely  for modelling quenches
 in  the Ising  SK  spin glass  as  it  is only  by  a full  dynamical
 treatment  that one  can  obtain  a $p(h)$  which  is  linear in  $h$
 \cite{eastham:06,horner:07,yan:15}.

In Sec.\ \ref{model} we introduce the KAS model. In Sec.\ \ref{sec:overlap} we investigate
how the quenched state depends on the initial state. In Sec.\ \ref{sec:marginality} we study the distribution of the local fields $p(h)$ of the quenched state, and from its form deduce that marginality only exists when $\sigma \le 1/2$.  The existence of SOC behavior is deduced from
a study of  the $N$ dependence of the spin glass susceptibility and the energy of the quenched state in Sec.\ \ref{sec:soc}.

 \begin{figure*}
  \includegraphics[width=0.5\columnwidth]{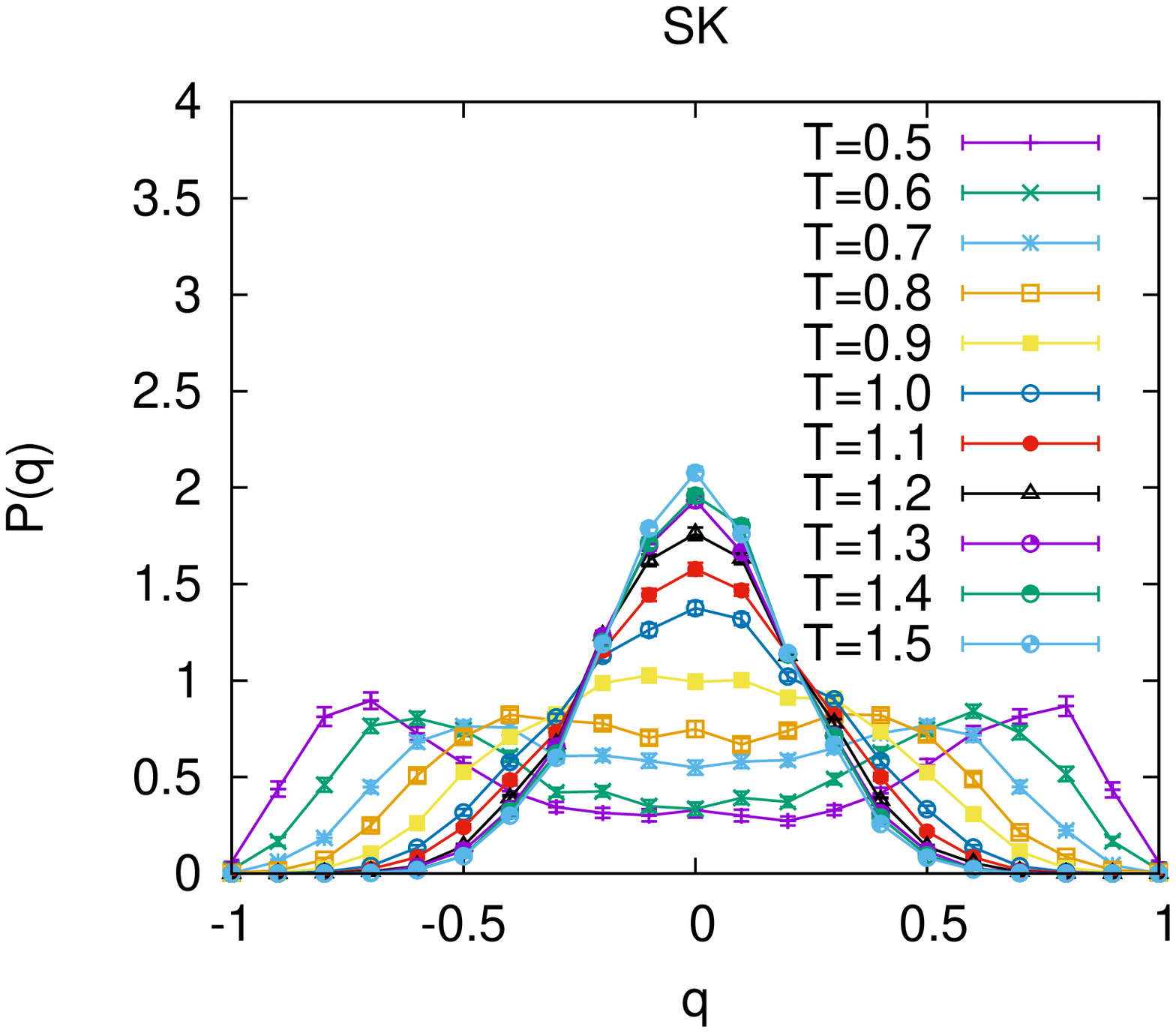} 
  \includegraphics[width=0.5\columnwidth]{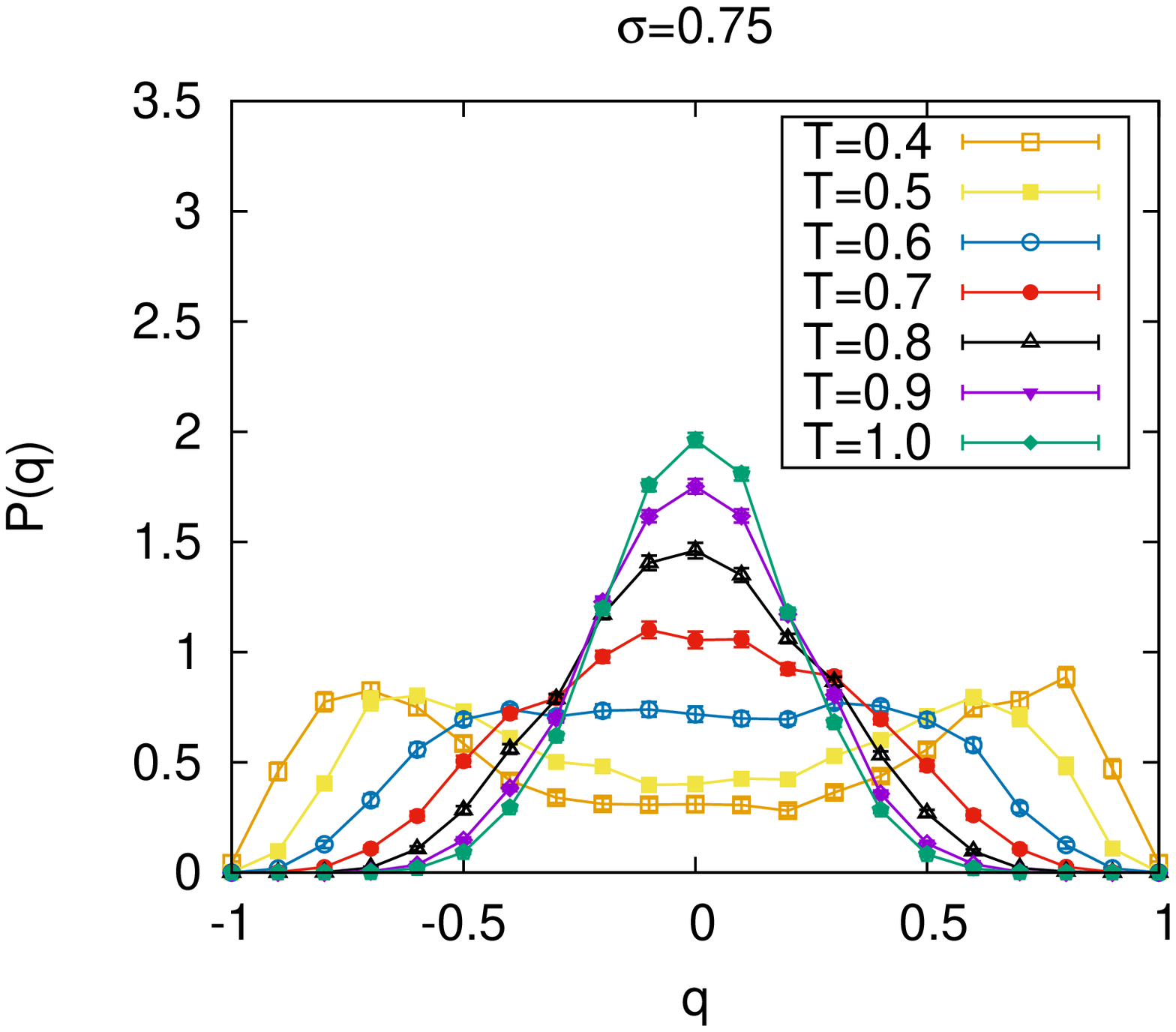} 
  \includegraphics[width=0.5\columnwidth]{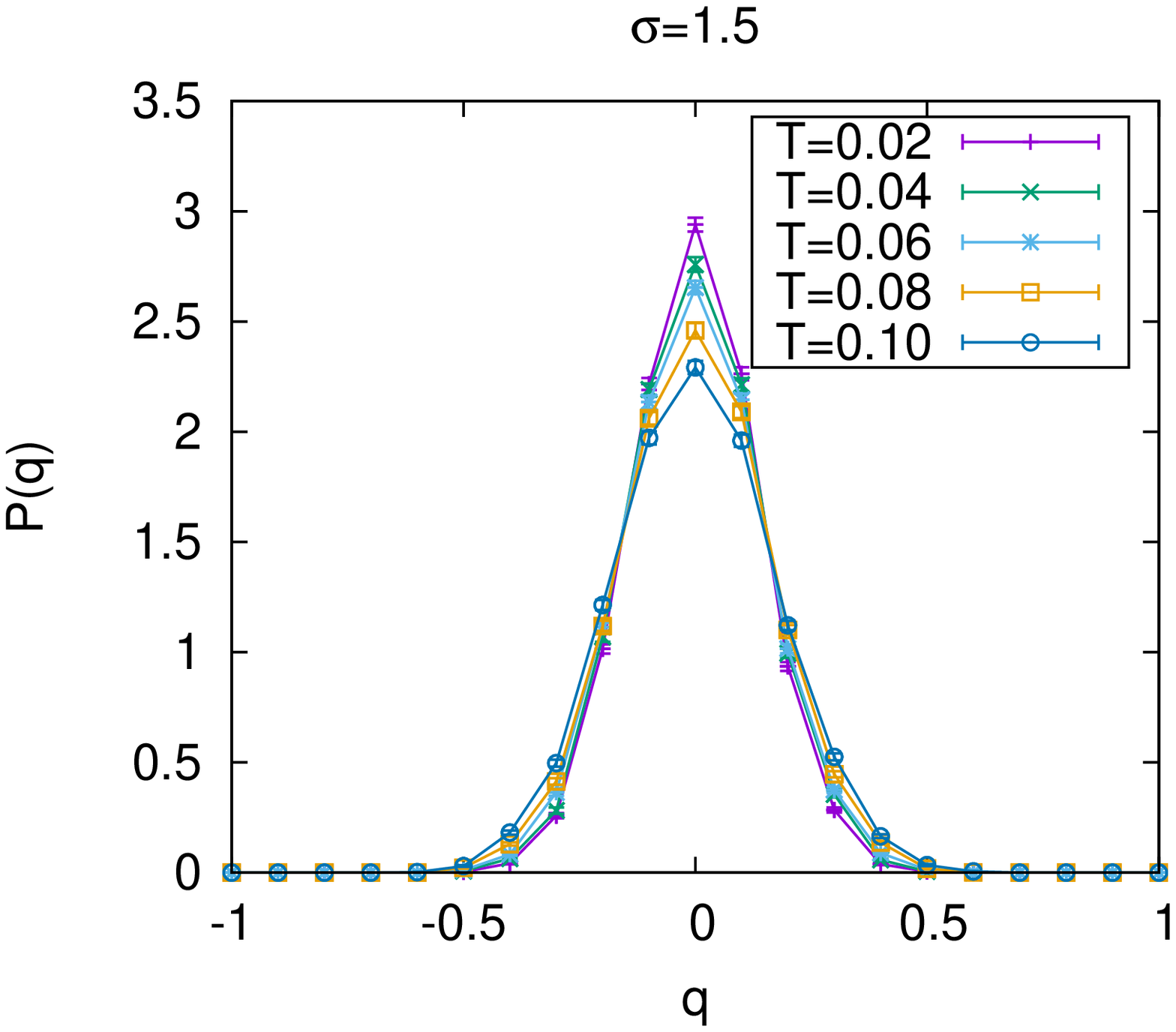} 
  \caption{The distribution of overlaps $P(q)$ of the states reached when the initial state is equilibrated at a temperature $T$. A range of temperatures $T$ is shown for three values of $\sigma = 0, 0.75, 1.50$. The system size is $N=256$. For each sample of disorder, $N_{min} = 15 $ minima are found and overlapped with each other, thus obtaining 105 different values of q from which an histogram $P(q)$  is obtained. This $P(q)$  is then averaged over $N_{samp} = 200$ samples of disorder, thus also extracting error-bars. 
The SK limit has a phase transition at $T_{c} = 1$, while at  $\sigma = 0.75$,  $T_{c} = 0.62$~\cite{ky:2003}. No  phase transition exists for $\sigma = 1.5$. Whether the final state displays a non-trivial $P(q)$ depends on whether the initial temperature $T$ is less than $T_{c}$ or not. For $\sigma=1.5$, a trivial $P(q)$, that is one approaching $\delta(q)$ as $N \to \infty$, is obtained even for quenches from  very low temperatures.}
  \label{figPQ}
\end{figure*}

\section{The Model}
\label{model}
 The Kotliar, Anderson and Stein (KAS) \cite{KAS} Hamiltonian is 
\begin{equation}
\mathcal{H} = -\sum_{\langle i, j \rangle} J_{ij} S_i S_j,
\label{Ham}
\end{equation}
where the Ising spins $S_i$ ($i = 1, 2, \cdots, N$), taking values $\pm 1$, are arranged in a circle of perimeter $N$. 
The geometric distance between sites $i$ and $j$ is $r_{ij}=\frac{N}{\pi}\sin\left[\frac{\pi}{N}(i-j)\right]$, the length of the chord between the sites $i,j$. The interactions  $J_{ij}$ are long-ranged and depend on the distance $r_{ij}$ as $J_{ij}=c(\sigma,N) \varepsilon_{ij}/r_{ij}^{\sigma}$, where $\varepsilon_{ij}$ is a Gaussian random variable of mean zero and unit variance. The coefficient $c(\sigma,N)$ is chosen to make the mean-field transition temperature $T_c^{{\rm MF}}$ equal to unity for all values of $\sigma$: 
\begin{equation}
  \left[ T^{{\rm MF}}_{{\rm SG}}(c) \right]^2 =\frac{1}{N} \sum_{i \not= j} \left[J_{ij}^{2}\right]_{{\rm av}}
  = c(\sigma,N)^{2} \sum_{j =2}^N\frac{1}{r^{2\sigma}_{1j}}=1. \label{Eq:T_SG_1d_pow_fullConnect}
\end{equation}
 The sum over $j$ can be done for large $N$ and gives
\begin{equation}
\frac{1}{c(\sigma, N)^{2}}=2 \zeta[2 \sigma]+\frac{\Gamma[1/2-\sigma]^2}{2^{2 \sigma} \Gamma[1-2 \sigma]}
\Big(\frac{N}{\pi}\Big)^{1-2 \sigma} +{\rm O}(\frac{1}{N^2}).
\label{cfix}
\end{equation}
Thus when $\sigma < 1/2$, $c(\sigma, N) \sim [1/N^{1/2-\sigma}](1+{\rm O}(1/N^{1-2 \sigma}))$, while for $\sigma > 1/2$, $c(\sigma,N) \sim (1+{\rm O}(1/N^{2 \sigma-1}))$. We have found when studying the energy per spin  $E$ reached in the quench that there is a finite size correction for $\sigma > 1/2$ of ${\rm O}(1/N^{2\sigma-1})$, whose origin is that $E \propto c(\sigma,N)$. Such corrections to scaling are large, especially when $\sigma$ is close to $1/2$. Note that this correction to scaling is associated  with the zero temperature fixed point, rather than the critical fixed point. It is only the discovery of this form for the leading correction to scaling that has enabled us to analyze our data. 

There is a mapping between $\sigma$ and an effective dimensionality $d_{\textrm{eff}}$ of the EA model \cite{ky:2003,katzgraber:09b,leuzzi:09,banos:12b,aspelmeier:16}. For $1/2 < \sigma < 2/3$, it is $d_{\textrm{eff}}=2/(2 \sigma-1)$; thus $\sigma =2/3$ corresponds to an effective dimensionality of $6$.

We generated one spin flip stable states by a quenching procedure that involves
repeatedly flipping spins to orient them with their local fields, according to the
sequential (as opposed to the `greedy' or `polite')
algorithm. Previous work \cite{parisi:95} suggests that although a
difference in the energy of the final state can be seen based on
the precise algorithm employed, the \textit{nature} of the final state is
independent of the algorithm. In the 
sequential algorithm used here, sites are scanned sequentially from $1$ through
$N$, and at each of them the spin is aligned to its local field, thus
monotonically reducing the energy of the system. When a spin is flipped the local fields $h_i$ are immediately updated.  The protocol of
repeatedly aligning spins is carried out until  convergence
is obtained. The initial state was either a random spin state, which corresponds to infinite temperature, or one of the  spin configurations of an equilibrated system at temperature $T$.

\section{Dependence of the Parisi overlap $P(q)$ on the initial state}
\label{sec:overlap}

The overlap between two minima $A$ and $B$ obtained after a quench is defined as
\begin{equation}
q \equiv \frac{1}{N}\sum_{i}S_i^{A} S_i^{B}.
\end{equation}
Its distribution $P(q)$ contains crucial information about the nature of the final state reached by the quench. In Fig.~\ref{figPQ} we have plotted the sample averaged distribution function $P(q)$  for systems of $N =256$ spins obtained in quenches from various temperatures $T$. For a quench which starts at a temperature $T > T_c$, the resulting $P(q)$ is trivial, in that it reduces in the large $N$ limit to $P(q)=\delta(q)$ \cite{newman:99}. Such a form indicates that the states obtained in the quench are completely uncorrelated from each other. However, for quenches which start from a temperature $T< T_c$ a non-trivial $P(q)$ was found. One would expect that the $P(q)$ obtained from a quench starting from $T<T_c$ has the replica symmetry breaking or replica symmetry features expected for that $\sigma$ value, whatever that might be. For $\sigma >1$ $T_c$ is expected to be zero, and our results at $\sigma =1.5$, shown  in the third panel, are consistent with the final state always being that expected from a quench which starts in the paramagnetic region.

\section{Marginality and the distribution of local fields $p(h)$}
\label{sec:marginality}

In this section we discuss the distribution of local fields $h_i$ after the quench from infinite temperature. The magnitude of $h_i$ after the quench is given by 
\begin{equation}
h_i = S_i \sum_j J_{ij} S_j,
\label{hidef}
\end{equation}
where $S_i$ are the spins at the end of the quench. Notice that $h_i>0$.  What will interest us mostly is whether the form of $p(h)$ provides any evidence for marginality, in the sense that the state which is reached is just on the edge of stability \cite{muller:15}. Our conclusion will be that the quenched state has indeed marginal stability if $\sigma \le 1/2$, but not if $\sigma >1/2$.

\begin{figure}
\includegraphics[width=0.8\columnwidth]{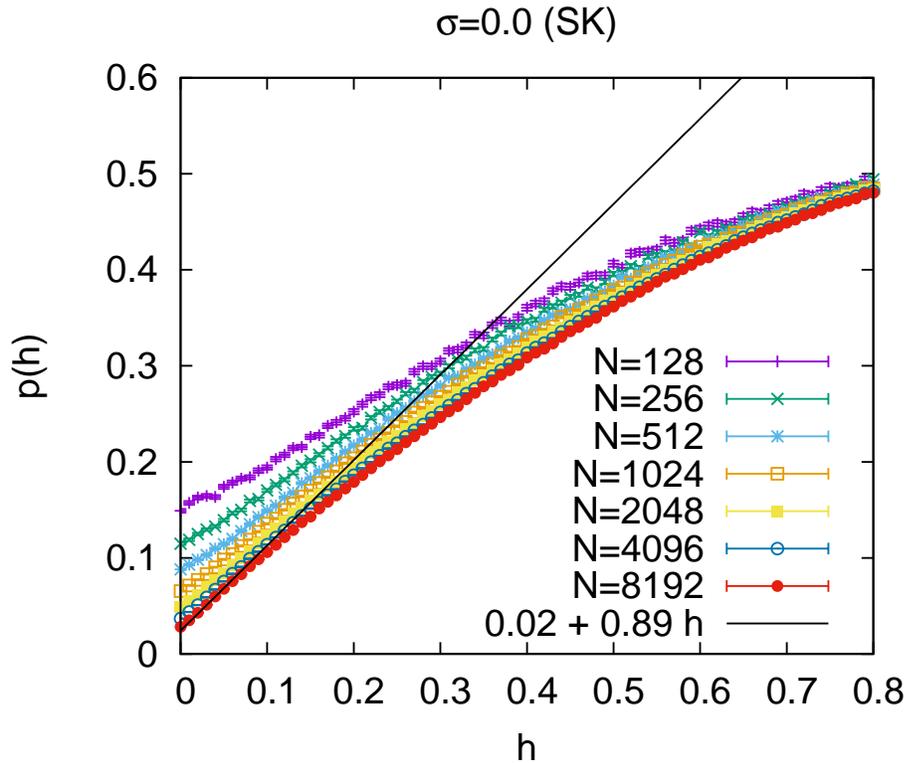} 
\caption{A plot of $p(h)$ after a quench from a random initial state, for $\sigma=0$ i.e.\ the SK model, for $h< 0.8$ for a range of $N$ values from 128 to 8192. In the large $N$ limit, $p(h) \sim 0.89 h$ at small $h$ if the system is marginal (using the value of $H$ predicted by  Eq.~(\ref{Hlt05}) with $p(h) =h/H^2$) and the black line is a line of that slope. Finite size effects cause $p(0)$ to be finite but this intercept on the $y$ axis decreases as $1/\sqrt{N}$.}
\label{SKphN}
\end{figure}

\begin{figure}
\includegraphics[width=0.8\columnwidth]{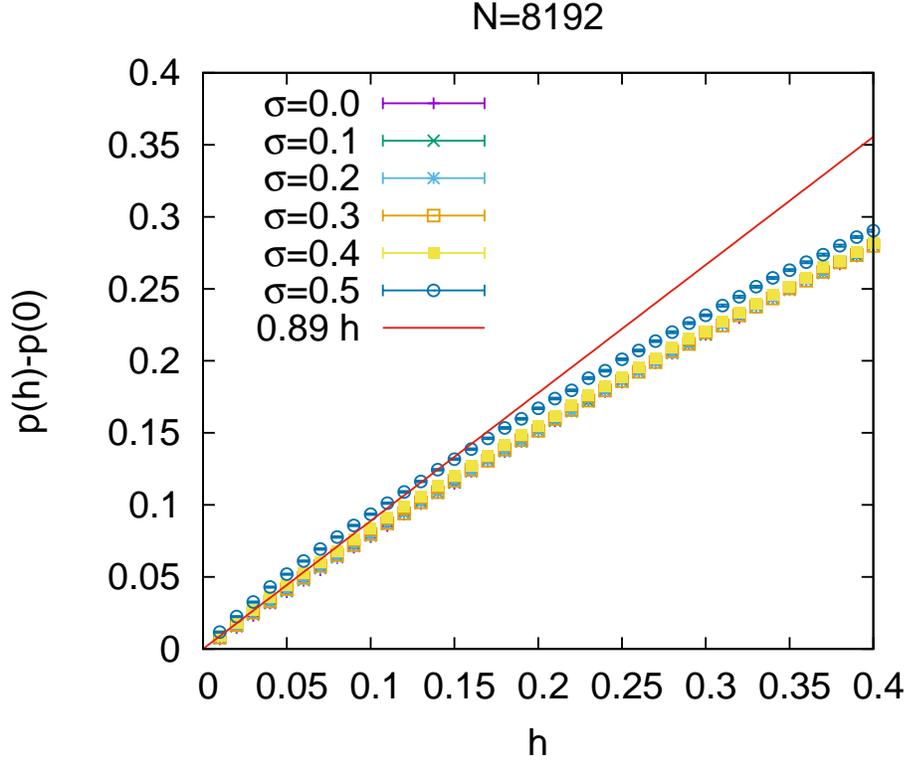} 
\caption{Plot of $p(h)-p(0)$ versus $h$ for values of $\sigma$, $0.0,0.1,0.2,0.3,0.4$, and $0.5$ for $N =8192$ after a quench from a random initial state. The subtraction of $p(0)$ is done to reduce finite size effects. The red line is a line of slope $0.89$ which is the value expected if the quenched states are just \textit{marginal}.}
\label{SKmarginality}
\end{figure}

We shall use the argument of P.\ W.\ Anderson (as reported in Ref.~\cite{palmer:79}) to obtain a ``bound'' on the local field distribution $p(h)$ for small $h$. For $\sigma < 1/2$, it is expected that $p(h)=h/H^2$ at small fields in the thermodynamic limit (see Fig.~\ref{SKphN}). In the state reached in the quench, relabel the sites in order of their increasing local field $h_i$ and consider the first $n$ of these sites, where $1 \ll n \ll N$. Suppose one flips all $n$ of the spins at these low-field sites: the consequent energy change is
\begin{equation}
\Delta E= 2 \sum_{i=1}^{n}h_i-2 \sum_{i=1}^{n}\sum_{j=1}^{n} J_{ij} S_i S_j.
\label{energyshift}
\end{equation}
$\Delta E$ would  be non-negative if the initial state were the ground state. Biroli and Monasson \cite{biroli:00} gave an argument  that for the SK model any state (and not just the ground state) is stable against flipping a finite number of spins. Their argument was that $\Delta E_{ij}= 2 h_i+2 h_j-2 J_{ij} S_iS_j$ which corresponds to flipping just two spins, reduces to $2h_i+2h_j$ in the large $N$ limit as $J_{ij}$ goes to zero as $1/N^{1/2}$ when $N \to \infty$ for the SK model. But $h_i$ and $h_j$ could themselves  be of order ${\rm O}(1/\sqrt{N})$ and excluding this possibility will give us a bound on the value of the coefficient $H$. 

The value of $h_n$ can be obtained from  solving
\begin{equation}
n =N\int_0^{h_n}\, dh\,\frac{h}{H^2}=\frac{Nh_n^2}{2 H^2}.
\label{nhn}
\end{equation}
The first term, $\Delta E_1$ in Eq.~(\ref{energyshift}) is similarly
\begin{equation}
\Delta E_1= 2 \sum_i^{n}h_i=2 N\int_0^{h_n}\, dh\,\frac{h^2}{H^2} 
 =\frac{2N h_n^3}{3 H^2}=\frac{4 \sqrt{2}}{3}H n \sqrt{\frac{n}{N}}.
\label{E1hn}
\end{equation}

For the case $\sigma>1/2$, it is expected that marginality \cite{muller:15} requires 
\begin{equation}
 p(h)= \frac{1}{H} \left(\frac{h}{H}\right)^{(1/\sigma)-1}.
\label{phdefgt05}
\end{equation}
 Eq.~(\ref{nhn}) becomes 
\begin{equation}
 n =\sigma N \left(\frac{h_n}{H}\right)^{1/\sigma},
\label{eqgt05}
\end{equation}
while  Eq.~(\ref{E1hn}) becomes 
\begin{equation}
\Delta E_1=2 n\left(\frac{n}{N}\right)^{\sigma} \frac {H}{(\sigma+1) \sigma^{\sigma}}.
\label{E1gt05}
\end{equation}

The second term in Eq.~(\ref{energyshift}) can be re-written as $\Delta E_2=-2 \sum_{i=1}^n S_i \Delta_i$, where $\Delta_i=\sum_{j=1}^n J_{ij} S_j$. $\Delta_i$ is a quantity which on average is zero.
 Its variance is
\begin{equation}
\frac 1 n \sum_{i=1}^n\Delta_i^2  =  \frac 1 n  \sum_{i=1}^n \sum_{j=1}^n J_{ij} S_j \sum_{k=1}^n J_{ik} S_k 
 =  \frac 1 n \sum_{i=1}^n \sum_{j=1}^n J_{ij}^2.
\label{Delta}
\end{equation}
 Suppose now that the positions of the spins $S_i$, $ i=1,2, \cdots, n$ are equally spaced so that $R_i =i N/n$, then the variance equals $(n^{2 \sigma}/N^{2\sigma}) c(\sigma,N)^2/c(\sigma,n)^2$ which reduces to $n/N$ when $\sigma <1/2$ and to $(n/N)^{2 \sigma}$ when $\sigma >1/2$.  ($c(\sigma,N)$ was defined in Eq.~(\ref{cfix}) and we have used its large $N$ and $n$ form). Since $S_i$ and $\Delta_i$ are correlated in sign, we  have for $\sigma <1/2$, $\Delta E_2 =-2 n \sqrt{n/N}$. For $\sigma > 1/2$, $\Delta E_2 =-2 n (n/N)^{\sigma}$.

Then the total energy $\Delta E=\Delta E_1 +\Delta E_2$
becomes for $\sigma <1/2$,
\begin{equation}
 \Delta E=2 n \left(\frac{n}{N}\right)^{1/2} \left[\frac{2\sqrt{2}}{3} H-1\right],
\label{Elt05}
\end{equation}
so the quenched state would be just marginal (i.e.\ has $\Delta E =0$) if
\begin{equation}
H= \frac{3}{2 \sqrt{2}}.
\label{Hlt05}
\end{equation}
For $\sigma >1/2$
\begin{equation}
\Delta E= 2 n \left(\frac {n}{N}\right)^{\sigma}\left[\frac{1}{(\sigma+1) \sigma ^{\sigma}} H-1\right].
\label{Eht05}
\end{equation}
Thus if the system is just marginal 
\begin{equation}
H=(\sigma+1) \sigma ^{\sigma}.
\label{Hgt05}
\end{equation}

\begin{figure}
\includegraphics[width=0.8\columnwidth]{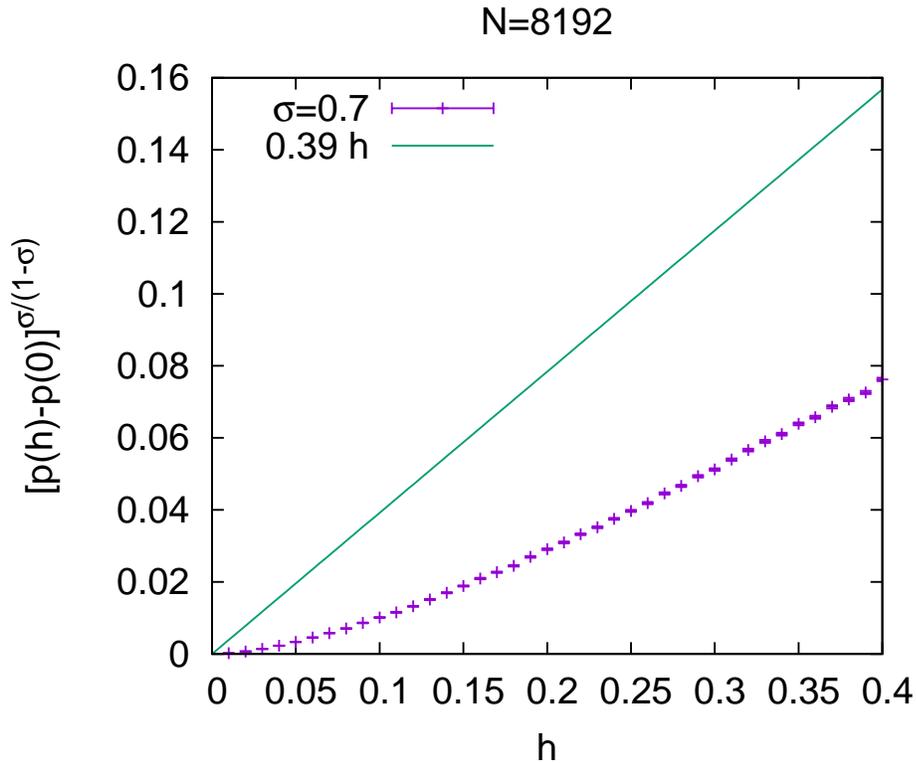}
\caption{$(p(h)-p(0))^{\sigma/(1-\sigma)}$ plotted versus $h$ for a quench from infinite temperature for a system of $N =8192$ spins at $\sigma =0.70$. The green line  corresponding to a straight line of the form $0.39 h$ is what would follow from Eq.~(\ref{phdefgt05}) with $H$ determined from Eq.~(\ref{Hgt05}). The agreement is very poor implying that marginality in the sense of the Anderson argument \cite{palmer:79} is not present.  }
\label{siggt05}
\end{figure}

\begin{figure}
  \includegraphics[width=0.8\columnwidth]{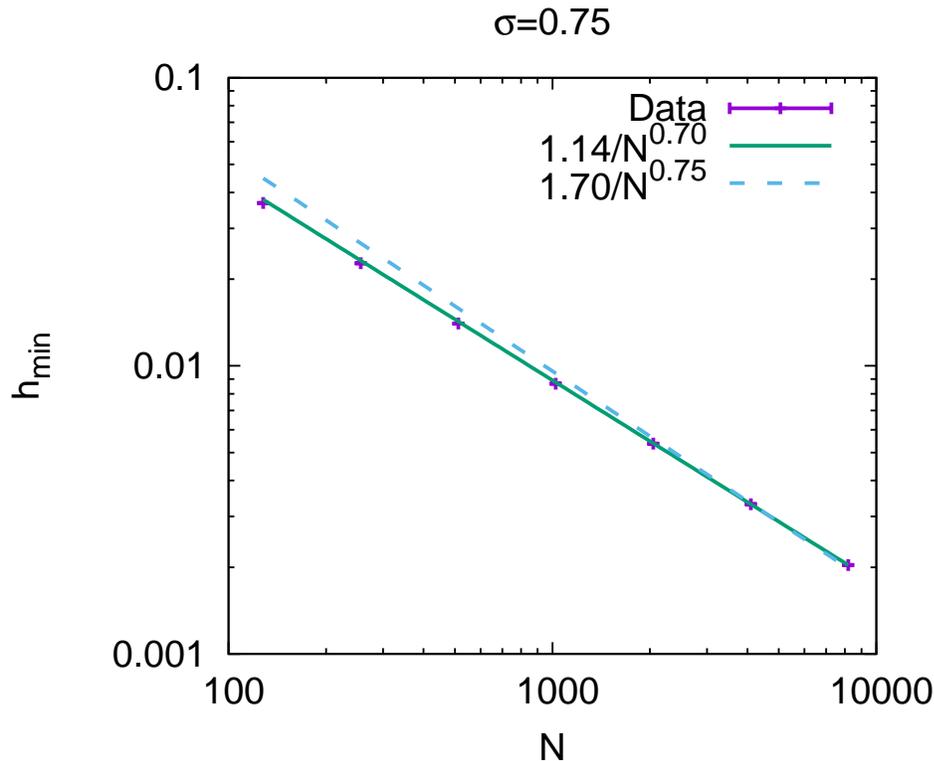} 
  \caption{Size dependence of the magnitude of the smallest local field for the special case of $\sigma = 0.75$. High statistics have been generated for this particular case in order to test for marginality. For system sizes, $N=128,\cdots,2048$, $16384$ samples of disorder have been generated, for the system size $N=4096$, we have $5246$ samples of disorder, while for the largest system size $N=8192$, we have $1258$ samples of disorder. A fit to the form $b/N^{c}$ fixing the value of $c=\sigma=0.75$ as per the expectation from marginality, is roughly consistent with the data from the largest system sizes, but a value for $c$ of $0.70$ fits the data better over the entire set of $N$ values studied.}
  \label{fighmin}
\end{figure}

\begin{figure*}
  \includegraphics[width=0.5\columnwidth]{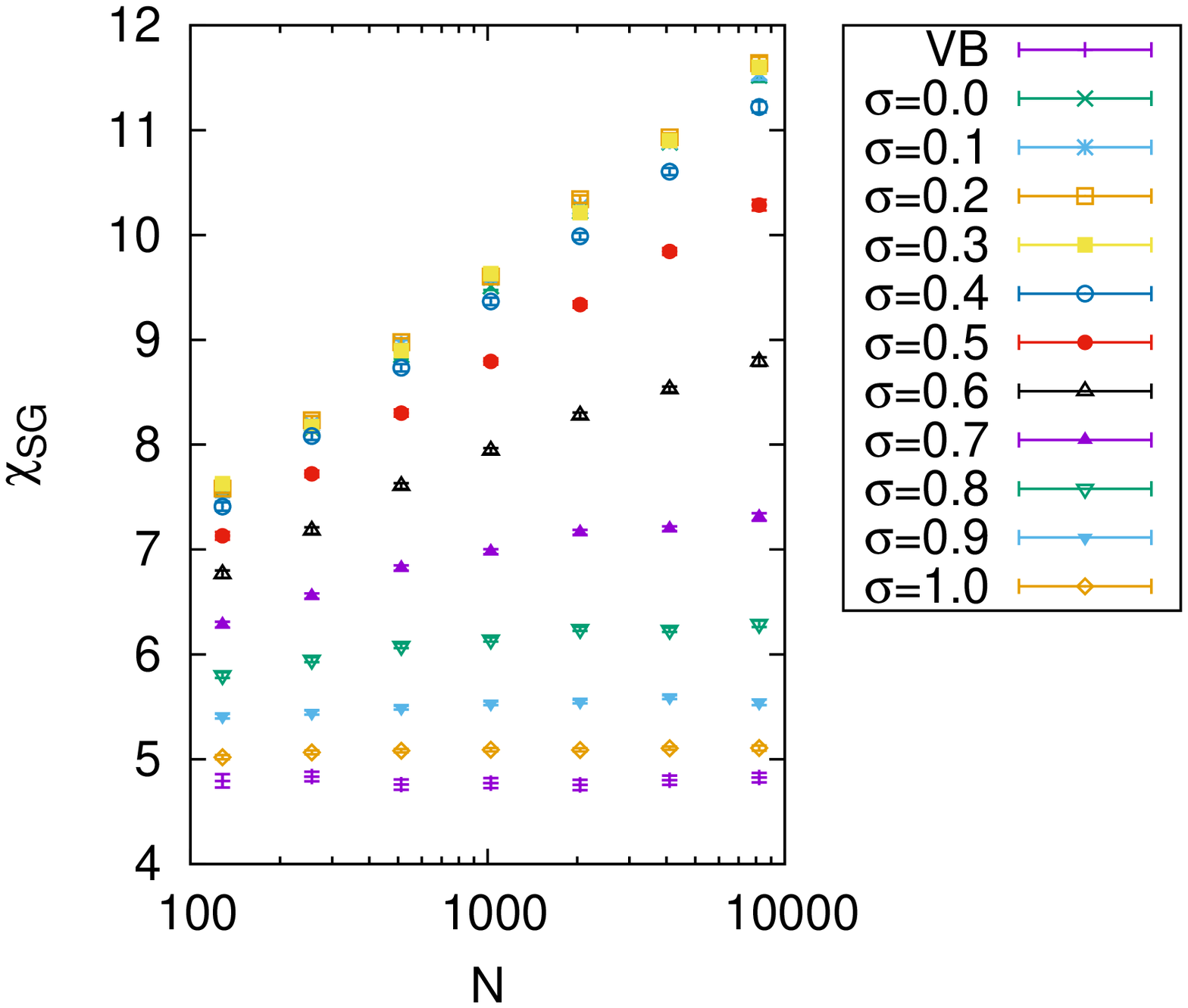} 
  \includegraphics[width=0.5\columnwidth]{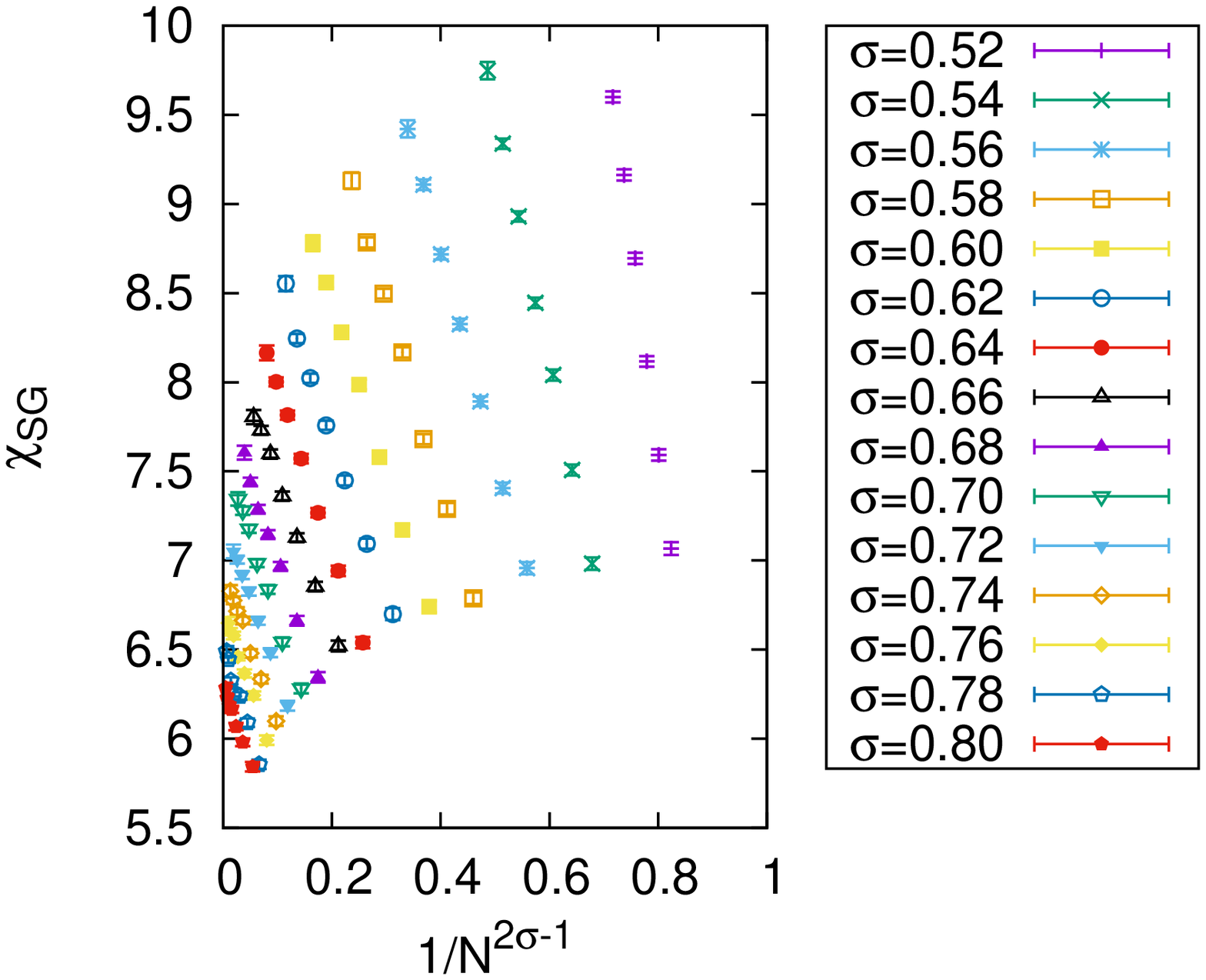} 
  \includegraphics[width=0.5\columnwidth]{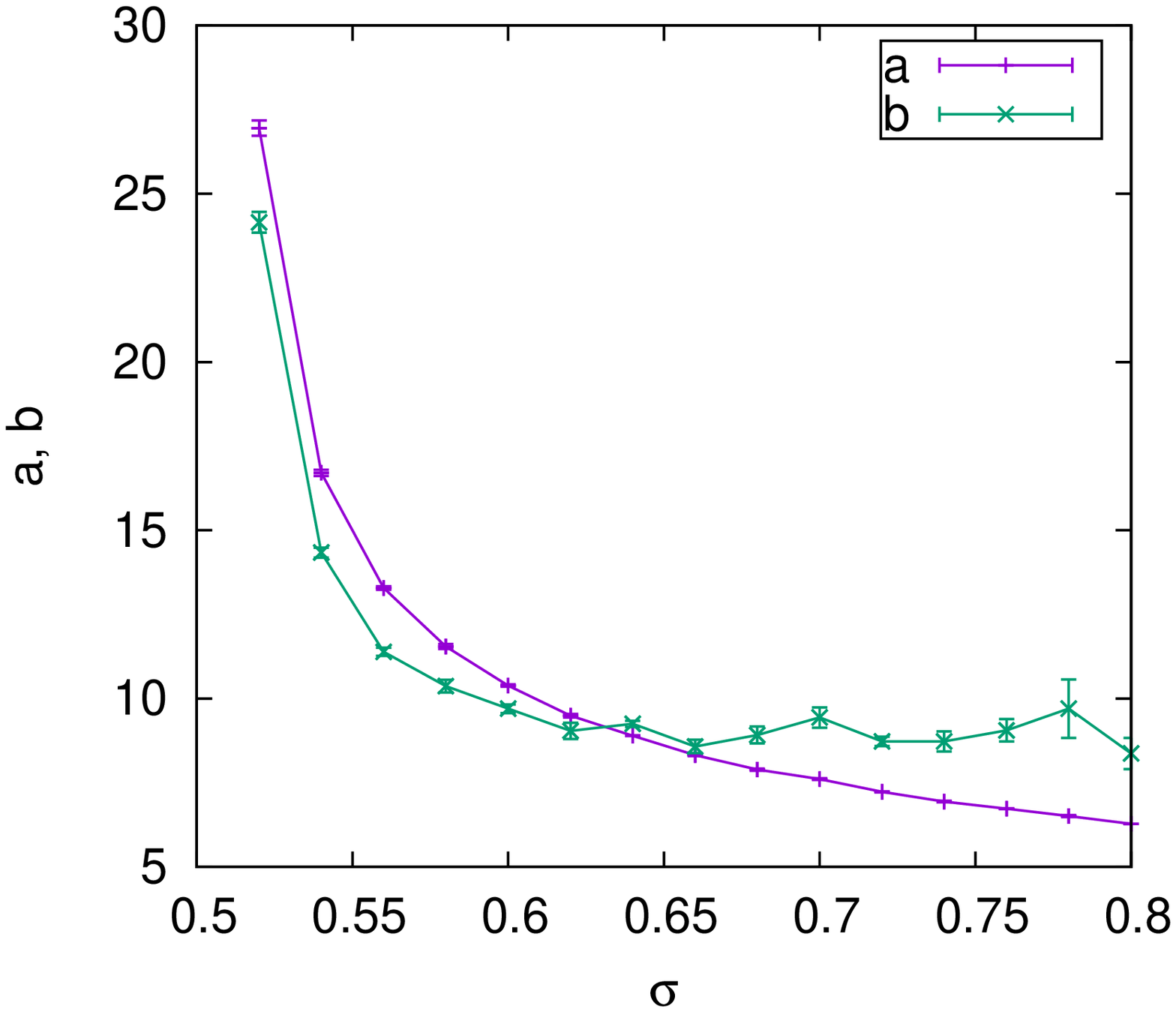} 
  \caption{The first two panels show the $N$-dependence of $\chi_{SG}$ as defined in Eq.~(\ref{chidef}) for a variety of $\sigma$. The number of samples used was $1000$ and the number of initial infinite temperature (random) 
configurations  was $15$. For $\sigma \le 0.5$, $\chi_{SG}$ diverges as a logarithm of the system-size. The data points for $\sigma=0.0,0.1,0.2,0.3$ are so close as to be barely distinguishable. For $\sigma >0.5$, there is a clear tendency for $\chi_{SG}$ to saturate at the largest sizes we are able to study: we are able to find very good fits to the saturating functional form $\chi_{SG} = a - b/N^{2\sigma-1}$ (see the second panel). The third panel shows the fit parameters $a,b$ as a function of $\sigma$. In the first panel VB refers to the Viana-Bray model which is a diluted version of the SK limit, $\sigma=0$, of the KAS model in which each spin is only coupled to six others: its size independent $\chi_{SG}$ shows that it lacks SOC.  Ref.~\cite{andresen:13} also reached the same conclusion based on their study of avalanches.}
  \label{figchisg}
\end{figure*}

\begin{figure*}
  \includegraphics[width=0.5\columnwidth]{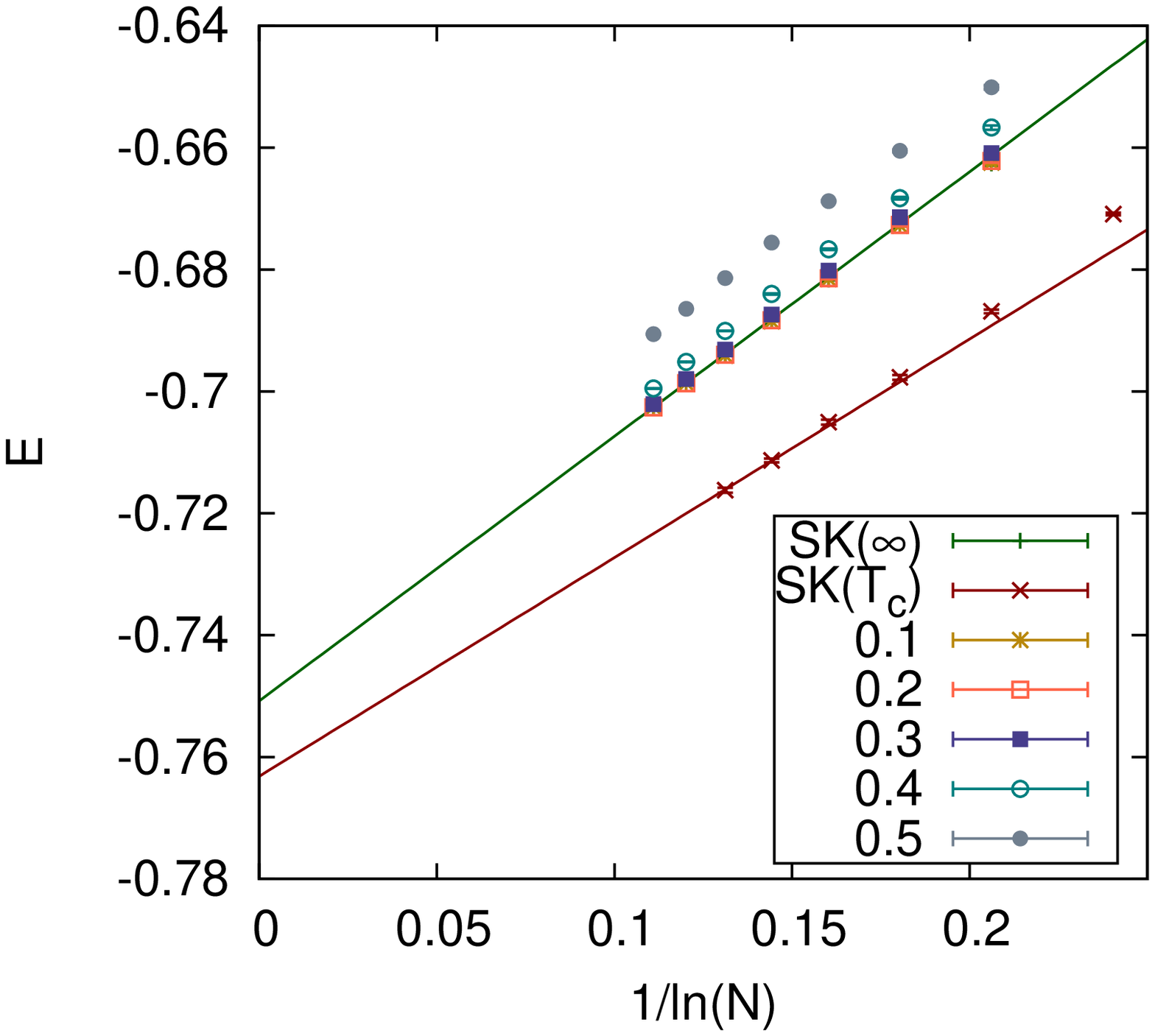} 
  \includegraphics[width=0.5\columnwidth]{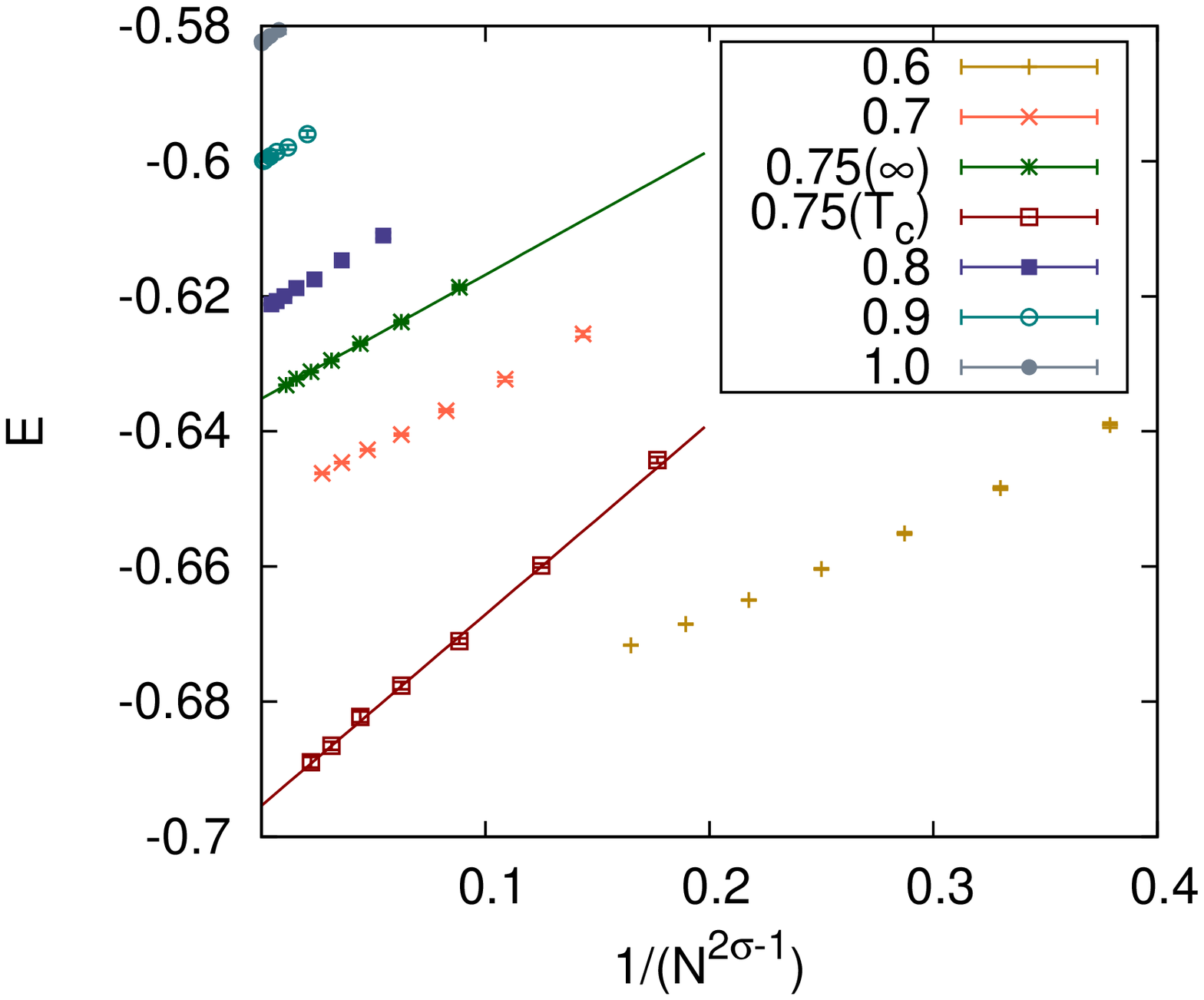} 
  \caption{Energy per spin $E$ as a function of system size $N$ after quenches, for $\sigma \le 0.5$ in the left panel and $\sigma >0.5$ in the right panel. In the quenches from the infinite temperature (random) initial state $1000$ samples were used with $15$ different initial starts. The quenches from $T_c$ were done only for the SK limit ($\sigma=0$ where $T_c=1$) and for  $\sigma =0.75$ where $T_c = 0.62$ \cite{ky:2003} and for these values of $\sigma$  straight lines have been drawn through the data points for both types of quench as a guide to the eye. For the quenches from $T_c$  only $200$ samples were averaged.  For $0\le\sigma\le 0.5$, the energy saturates to a characteristic energy $E_{c}$ as $1/\ln N$ for both the quench from infinite temperature and from $T_c$. For $\sigma >0.5$, the energy of quenches fits well to the form $c+d/N^{2\sigma-1}$.  }
  \label{figenergy}
\end{figure*}

In Fig.~\ref{SKmarginality} we have plotted $p(h)-p(0)$ (the subtraction of $p(0)$ is to reduce the consequences of the finite size intercept on the $y$-axis) as a function of $h$ for some $\sigma$ values less than $1/2$. The slope of the red line which is drawn using the value of $H$ which makes the system just marginal agrees quite well with the data for $\sigma < 1/2$. In Fig.~\ref{siggt05} we have repeated the exercise for $\sigma =0.7$. The green line is the line which would be expected if the system is just marginal. The data points are not close to this expectation at all and indicate that the state reached in the quench is stable  (i.e.\ $\Delta E >0$) by the Anderson criterion. We have examined other $\sigma$ values which are greater than $0.5$ and have found that the size of the discrepancy increases steadily as $\sigma$ rises above $0.5$.

Another confirmation that the quenched state  is not marginal for $\sigma >1/2$ is provided by Fig.~\ref{fighmin}. An assumption behind marginality is that Eq.~(\ref{phdefgt05}) should  hold. If that is the case, then in the large $N$ limit the smallest value of $h$, $h_{min}$ should decrease as $1/N^{\sigma}$ (see Eq.~(\ref{eqgt05}) with $n=1$).  The results for $\sigma =0.75$ in Fig.~\ref{fighmin} indicate that a better fit to the data is as $h_{min} \sim 1/N^{0.70}$. However, the discrepancy is modest for the exponent.

The states generated in the quench for $\sigma >1/2$ seem to be stable according to the Anderson argument, where one examines the stability against flipping the  spins in the first $n$ smallest fields, (see Fig. ~\ref{siggt05}), as $H$ is larger than the just marginal value $(\sigma+1) \sigma^{\sigma}$ if one determines it from the slope of the data at small $h$.  However, we suspect that they are unstable by the argument of Biroli and Monasson against flips of two spins where the fields  of the flipped spins are not restricted to be small as in the Anderson argument.  In other words, the states generated by the quench are just one spin flip stable states. In the SK limit a state generated by the quench will be a pure state \cite{biroli:00}, stable against flipping an arbitrary number of spins. 

In the next section we show that the existence of marginality for $\sigma \le1/2$ seems to be associated with self-organized criticality as we can only find that when $\sigma \le  1/2$.

\section{Self-organized criticality}
\label{sec:soc}

We have made  a finite size scaling study of the spin-glass susceptibility $\chi_{SG}$:
\begin{equation}
\chi_{SG} = \frac{1}{N}\sum_{i,j}\big[\langle S_{i}S_{j}\rangle^{2}\big]_{av},
\label{chidef}
\end{equation}
where the angular brackets represent an average over the metastable minima for a given sample of disorder. The minima in this case were obtained  from a random initial state, so that we are studying the case where $P(q)$ is trivial. Note that $ \chi_{SG}= N {\rm Variance}(q^2)$.

In the regime $0\le\sigma\le 1/2$, $\chi_{SG}$ appears to diverge as $\ln(N)$, whereas for the region $\sigma>1/2$, $\chi_{SG}$ saturates to a finite value at large $N$, the form of this dependence being well fitted by $\chi_{SG} = a-b/N^{2\sigma-1}$ (see Fig.~\ref{figchisg}).  The coefficients $a, b$ appear to approach each other and diverge as $\sigma \to 0.5^+$ (see Fig.~\ref{figchisg}). Note that
\begin{equation}
 \frac{1}{2\sigma-1}\left[1-\frac{1}{N^{2\sigma-1}}\right] \to \ln (N)
\label{limit}
\end{equation}
 in the limit $\sigma\to 0.5$. Thus the  divergence  of $\chi_{SG}$ for $\sigma \le 1/2$ as $\ln N$ seems to be natural if one takes the Mori argument~\cite{mori:11,wittmann:12} that all systems for $\sigma\le 1/2$ behave in the same way, and just as in the SK limit of $\sigma =0$. Consistent with this finding for $\chi_{SG}$, Fig.~\ref{figenergy} shows that the energy reached by the quench $E(N)$ goes as $E_{c}+{\rm const}/\ln(N)$ for $\sigma<0.5$, almost independent of $\sigma$, at least for $\sigma =0.0, 0.1, 0.2$ and $0.3$: only the data points for $\sigma =0.4$ and $0.5$ differ significantly and for them the finite size corrections are  very large.   The Mori argument says that in the thermodynamic limit quantities such as the energy should be independent of the value of $\sigma$ when it is less than $0.5$. However,  for $\sigma > 1/2$ the energy  $E(N)$ behaves quite differently and the right panel of Fig.~ \ref{figenergy} shows that it goes as $c+d/N^{2\sigma-1}$, just as could have been anticipated from the $N$ dependence of $c(\sigma,N)$.

We next explain why these results are consistent with SOC behavior for $\sigma  \le 1/2$.  There is an energy $E_c$  in the large $N$ limit which separates  minima  which are just at the brink of having a non-trivial form for $P(q)$ from those at higher energy which have trivial overlaps. This has been established for the SK model when the average is taken over all one spin-flip stable states \cite{bm:81a}.  $E_c$ marks the transition to a state with broken replica symmetry. We expect that there will be a similar critical energy for states prepared by quenches from infinite temperature, and that its numerical value will depend on the quench procedure. 

At $E_c$ massless modes are present: that is near $E_c$ the system has marginal stability. We learned in Sec.\ III that for $\sigma <1/2$ the state reached in the quench had marginal stability so we would expect that the energy reached in the quench is close to $E_c$. $E_c$ is the analogue of the transition temperature $T_c$ in studies of the thermal spin glass susceptibility \cite{bm:81a} as the spin glass susceptibility diverges for $\sigma  \le  1/2$ as $\chi_{SG}  \sim 1/\tau$, the usual mean-field form, where $\tau=(1-T_c/T)$. Our quenches take us close to $E_c$ but miss by an amount of ${\rm O}(1/\ln N)$ due to finite size effects; our analogue of $\tau$ is $\sim 1/\ln N$, so $\chi_{SG} \sim \ln N$. This result is also consistent with our 
argument by continuity from $\sigma> 1/2$  in Eq.~(\ref{limit}).

For quenches from a temperature $T> T_c$ one would expect that the extrapolated energy $E_c(T)$ would be slightly different from that obtained in  the quench from infinite temperature.  Fig.~\ref{figenergy} shows that the quench from $T_c$ for the SK model goes indeed  to a somewhat lower value of the energy by an amount  ($\sim 1\%$) from that at infinite temperature, (and which is incidentally very close to the Parisi type estimates of the true ground state per spin: $E_g=-0.763 166 772 65(6)...$ \cite{oppermann:07}).
 The quench from $T_c$ is from an initial state where there are long-range correlations  which are absent for the initial state at infinite temperature, which is probably why their associated values of $E_c$ differ.
 
 For $ \sigma >1/2$ we did not see marginal stability and so the energy reached in the quenched state is probably not close to any critical value below which replica symmetry breaking effects might become visible. For $\sigma > 2/3$ we doubt  even the existence of any states with broken replica symmetry \cite{moore:11}.  Because of the absence of marginality it is no surprise really that $\chi_{SG}$ shows no sign of diverging with increasing $N$.  Figure \ref{figchisg} indicates that it is approaching a finite value as $1/N^{2 \sigma-1}$.
 
 We suspect that the existence of SOC behavior and marginality only for $\sigma \le 1/2$ might be reflected by differences in the nature of spin avalanches for $\sigma$ above and below $1/2$. Horner \cite{horner:07} found that  in the SK model  the number of spin flips per site before the final quenched state was reached increased $\approx \ln N$. However, Andresen and others  \cite{andresen:13,zhu:14} found  avalanches on the scale of the system size $N$ only when $z$, the number of neighbors of a given site  increases with $N$, as happens in the SK model. We would imagine such behavior would extend up to $\sigma =1/2$. For $\sigma > 1/2$ the effective number of neighbors is  finite (even though the critical behavior remains mean-field like up to $\sigma =2/3$; the KAS model with $1/2 <\sigma <2/3$ maps to the nearest-neighbor Edwards-Anderson model in a  dimension $d> 6$). Thus the dynamics would be expected to change at $\sigma =1/2$ along with the disappearance of SOC behavior and marginality.
 
 \section{Conclusions}
 
In the KAS model we have discovered that there is a connection  between SOC behavior  and marginality. Because the states reached in the quenches are marginal when $\sigma \le 1/2$, they are near the energy at which the states  have massless modes i.e.\ are becoming critical. In this case, the criticality is that associated with the onset of replica symmetry breaking.

It would be interesting to know whether in the many systems which are thought to have marginal behavior \cite{muller:15}, there is a similar connection with self-organized critical behavior. What is striking about the KAS model is that the transition which is self-organized can be identified;  it is the transition to states with correlations between them due to the onset of broken replica symmetry.
 
\section*{Acknowledgments} 
We would like to thank Markus M\"{u}ller for helpful discussions and Juan Carlos Andresen for an initial study of  $P(q)$.
AS acknowledges support from the DST-INSPIRE Faculty Award [DST/INSPIRE/04/2014/002461]. Some of the simulations in this project were run on the High Performance Computing (HPC) facility at IISER Bhopal. JY was supported by
Basic Science Research Program through the National Research Foundation 
of Korea (NRF) funded by the Ministry of Education (NRF-2017R1D1A09000527).

\section*{References} 
\bibliography{refs}

\end{document}